%% file: main.tex
\documentclass[conference]{IEEEtran}
\IEEEoverridecommandlockouts

\usepackage{cite}
\usepackage{amsmath,amssymb,amsfonts}
\usepackage{algorithmic}
\usepackage{graphicx}
\usepackage{textcomp}
\usepackage{xcolor}
\usepackage{comment}
\usepackage{xurl} 
\usepackage{caption}
\usepackage{subcaption}
\usepackage{bd}
\usepackage{hyperref}

\newcommand{\evsee}{\textit{EVSEE}}

\begin{document}

\title{Beyond Uptime: Actionable Performance Metrics for EV Charging Site Operators}

\author{\IEEEauthorblockN{Brandon D'Agostino}
\IEEEauthorblockA{\textit{Department of Electrical Engineering} \\
\textit{Stanford University}\\
Stanford, USA \\
bdag@stanford.edu}
\and
\IEEEauthorblockN{Jimmy Chen}
\IEEEauthorblockA{\textit{Precourt Institute for Energy} \\
\textit{Stanford University}\\
Stanford, USA \\
jimchen@stanford.edu}
\and
\IEEEauthorblockN{Ram Rajagopal}
\IEEEauthorblockA{\textit{Department of Electrical Engineering} \\
\textit{Department of Civil and Environmental Engineering} \\
\textit{Stanford University}\\
Stanford, USA \\
ramr@stanford.edu}
}

\maketitle

\begin{abstract}
The transition to electric vehicles (EVs) depends heavily on the reliability of charging infrastructure, yet approximately 1 in 5 drivers report being unable to charge during station visits due to inoperable equipment. While regulatory efforts such as the National Electric Vehicle Infrastructure (NEVI) program have established uptime requirements, these metrics are often simplistic, delayed, and fail to provide the diagnostic granularity needed by Charging Site Operators (CSOs). Despite their pivotal role in maintaining and improving site performance, CSOs have been largely overlooked by existing reporting standards. 

In this paper, we propose a suite of readily computable, actionable performance metrics---Fault Time, Fault-Reason Time, and Unreachable Time---that decompose charger behavior into operationally meaningful states. Unlike traditional uptime, these metrics are defined over configurable periods and distinguish between hardware malfunctions and network connectivity issues. We demonstrate the implementation of these metrics via an open-source tool that derives performance data from existing infrastructure without requiring hardware modifications. A case study involving 98 chargers at a California academic institution spanning 2018–2024 demonstrates that these metrics reveal persistent "zombie chargers" and high-frequency network instability that remain hidden in standard annual reporting.
\end{abstract}


\input{sections/1-introduction.tex}
\input{sections/2-background.tex}
\input{sections/3-site-operators.tex}
\input{sections/4-new-metrics.tex}

\input{sections/5-case-study.tex}
\input{sections/6-conclusion.tex}

\bibliographystyle{IEEEtran}
\bibliography{references.bib}

\appendices
\input{sections/A-evsee.tex}
\input{sections/B-integration-plugin}

\end{document}

%% file: sections/1-introduction.tex
\section{Introduction}
\label{sec:introduction}

The transportation sector is a substantial contributor to global greenhouse gas emissions, accounting for nearly 30\% of
emissions in the United States alone~\cite{USTransportationSector2022}. Alternative fuel vehicles have meanwhile
appeared as a promising solution to reducing emissions within this sector, with battery and plug-in hybrid electric
vehicles (EVs) emerging as a leading contender for clean light-duty transportation. EVs are thus experiencing increasing
market share and policy support worldwide, with the global EV fleet estimated to grow twelve-fold to encompass 525
million vehicles by 2035~\cite{ritchieTrackingGlobalData2024,GlobalEVOutlook2024a}.

\emph{EV charging} (i.e., refueling) is an integral component of both EV operations and adoption, and must therefore
work reliably. However, a growing body of evidence suggests this may not be the case in the U.S.
From 2021--2024, the consumer intelligence firm J.D. Power found that
19--24\%---or approximately 1 in 5---of EV drivers visited a charging station but were unable to charge their vehicle,
citing out-of-service or inoperable chargers as the prevailing
culprit~\cite{2024USElectric2024}.
Similarly, a 2022 study conducted in the San Francisco Bay Area found that 23.5\% of the surveyed charging ports in the area were
nonfunctional~\cite{rempelReliabilityOpenPublic2023}. Several news articles have subsequently highlighted these and
similar findings as well as the broader implications of unreliable charging on EV
adoption~\cite{hannahlutzEVDriversStruggle2023,jeffst.johnEVChargersHave2023,jeffst.johnWhatsEpidemicUnreliable2023,osakaHeresBiggestHurdle2023,rosevearEVChargingNeeds2023,davidrbakerWhyManyElectric2023,russmitchellPublicEVChargers2023}.

Despite an apparent consensus on the prevalence and severity of these issues, insufficient charging performance
reporting standards and limited performance data have made them challenging to diagnose and
address~\cite{keithBuildingSustainingReliable2022,CSPRFrameworkTechnical2023,jeffst.johnEVChargersHave2023}. Responses
to these challenges have emerged primarily on two complementary fronts: \emph{i)} the development of new standards and
recommendations for charging performance and reporting~\cite{CSPRFrameworkTechnical2023,
WhatNewOCPP2023,christinjeffersNRELSupportsEfforts2023,caseyquinnCustomerFocusedKeyPerformance2024,
mayureshsavargaonkarImplementationGuideCustomerFocused2024,MinimumRequiredError2023,mariaeduardamontezzocoelhoRecommendationsMinimumRequired2025,moriartyBestPracticesPayment2024};
and \emph{ii)} regulatory efforts to enforce these standards and
recommendations~\cite{NationalElectricVehicle,NationalElectricVehicle2023,adamdavisAssemblyBill21272024}. However, these
efforts are predominantly future oriented and do not directly address the ongoing performance issues at existing
charging sites around the country. Moreover, these efforts fall short in meeting the needs of \emph{charging site operators} (CSOs, or site operators), who play a crucial role in assessing, validating, and improving charging performance and reliability.

In this paper, we highlight that site operators have been largely overlooked by existing efforts to improve charging performance despite their pivotal role in doing so. Specifically, we scrutinize site operators' limited access to useful performance data and how existing, non-actionable performance metrics fail to meet their needs. In response, we propose a set of readily computable performance metrics for site operators
that provide the actionable insights they need to drive performance improvements at their sites.
Finally, we present a case study  
to demonstrate that our proposed metrics are not only readily computable at an existing charging site without
changes to the existing infrastructure, but capable of deriving actionable intelligence at the diagnostic granularity required for site operators to effectively assess, validate, and ultimately address performance issues.

In summary, this paper makes the following contributions:
\begin{itemize}
\item We highlight the importance of site operators in driving improvements to EV charging performance in the U.S., and
how they have been overlooked by existing charging performance reporting and improvement efforts.
\item We describe how existing charging performance metrics fall short in addressing site operators' needs, and propose
improved performance metrics that are readily interpretable by site operators and provide actionable insights to drive
performance improvements.
\item We demonstrate that these proposed metrics can be readily computed at existing charging sites using our, and present the performance insights they reveal through a case study at
a charging site hosted by an academic academic institution in California.
\end{itemize}


%% file: sections/2-background.tex
\section{Background}\label{sec:background}

Works highlighting EV charging reliability issues in the U.S. have emerged across industry, academia, and the media.
From 2021--2024, the consumer intelligence firm J.D. Power found that 19--24\% of EV drivers visited a public charging
station but were unable to charge their vehicle~\cite{2024USElectric2024},
with the 2024 report stating the predominant problem nationwide (responsible for 61\% of the 19\% non-charge
visits experienced by EV owners) to be that chargers were out of service or ``just wouldn't
work''~\cite{2024USElectric2024}. Similarly, a 2022 study conducted in the San Francisco Bay Area found that 23.5\% of
the 655 public combined charging system (CCS) DC fast charger (DCFC) ports surveyed were nonfunctional due to broken
connectors (0.6\%), blank or nonresponsive screens (3.7\%), on-screen error messages (4.6\%), network or connection
errors (0.9\%), payment system failures (7.6\%), and charge initiation failures
(6.1\%)~\cite{rempelReliabilityOpenPublic2023}. 
Several news articles have subsequently underscored these and similar findings as well as the deleterious impact
of unreliable charging on EV
adoption~\cite{hannahlutzEVDriversStruggle2023,jeffst.johnEVChargersHave2023,jeffst.johnWhatsEpidemicUnreliable2023,osakaHeresBiggestHurdle2023,rosevearEVChargingNeeds2023,davidrbakerWhyManyElectric2023,russmitchellPublicEVChargers2023}.

Many works have further highlighted the complexity of the charging ecosystem, limited availability of charging data, and
insufficient charging performance reporting standards and regulation as significant challenges to diagnosing and
addressing these
issues~\cite{keithBuildingSustainingReliable2022,CSPRFrameworkTechnical2023,jeffst.johnEVChargersHave2023}.

Initial responses to these challenges emerged primarily on the recommendations and technical standards front. In early
2023, SAE International---in collaboration with several automakers, EV charger manufacturers, charging network
providers, and other prominent stakeholders in the EV charging ecosystem---released the Charging System Performance
Reporting (CSPR) Framework~\cite{CSPRFrameworkTechnical2023}. This technical report identified charging reliability and
the lack of standardized means to gauge the performance of charging systems as crucial issues to address, and
contributed \emph{i)} the finding that existing charging system protocols could not fully meet the goals and
requirements of performance reporting; and \emph{ii)} recommendations for standardized performance metrics and
communication protocols for charging systems. Since the publication of the CSPR Framework, the latest revisions (i.e.,
versions 2.0.1 and 2.1) of the Open Charge Point Protocol (OCPP)---the predominant standard for communication between EV
chargers and central management systems---include significant enhancements to charging station monitoring and
control~\cite{CSPRFrameworkTechnical2023, WhatNewOCPP2023}. Later in 2023, the National Charging Experience Consortium
(ChargeX Consortium)---a joint effort between several national laboratories and a similar cross section of the EV
charging community---was established to address major charging experience
challenges~\cite{christinjeffersNRELSupportsEfforts2023}, and has since produced several recommendations on
customer-focused key charging performance indicators~\cite{caseyquinnCustomerFocusedKeyPerformance2024,
mayureshsavargaonkarImplementationGuideCustomerFocused2024}, minimum-required error
codes~\cite{MinimumRequiredError2023}, diagnostic
information~\cite{mariaeduardamontezzocoelhoRecommendationsMinimumRequired2025}, and payment
systems~\cite{moriartyBestPracticesPayment2024}.

Complementary efforts have also emerged on the regulatory front. The Infrastructure Investment and Jobs Act (IIJA; also
known as the Bipartisan Infrastructure Law, or BIL) enacted in 2021 established the National Electric Vehicle
Infrastructure Formula Program (NEVI Formula), a US\$5 billion program administered by the U.S. Department of
Transportation Federal Highway Administration (FHWA) to help states strategically deploy EV chargers. The final rule
released in 2023 requires all NEVI Formula-funded charging installations to adopt OCPP 2.0.1 as well as adhere to the
program's prescribed charger uptime and performance reporting
requirements~\cite{NationalElectricVehicle,NationalElectricVehicle2023}. 
Some states have also followed suit with their own legislation.
California, for instance, has enacted legislation that requires the California Energy Commission (CEC) to \textit{i)}
biennially assess the uptime of all chargers statewide; \textit{ii)} develop uptime recordkeeping and reporting
standards; and \textit{iii)} set uptime, operations, and maintenance requirements for publicly and ratepayer-funded
chargers~\cite{adamdavisAssemblyBill21272024}.

While these efforts are undoubtedly promising steps in the right direction, they are not without their own limitations.
For one, these efforts are predominantly future oriented and do not readily address the ongoing performance issues at
existing charging sites; rather, they focus on establishing new standards and guidelines that may not be applicable to
older infrastructure. Furthermore, without adoption mandates, these standards and guidelines may not be widely
implemented. For example, to the best of our knowledge, no adoption mandates for the CSPR Framework or ChargeX
Consortium recommendations currently exist. Moreover, industry experts and government officials have said that the NEVI
Formula's uptime requirements are ``relatively simplistic'' and ``necessary but not
sufficient''~\cite{jeffst.johnWhatsEpidemicUnreliable2023} to ensure charging reliability, and that insufficient
availability of or access to relevant charging performance data remains a
challenge~\cite{jeffst.johnWhatsEpidemicUnreliable2023}.

%% file: sections/3-site-operators.tex
\section{The Overlooked Role of Site Operators}\label{sec:site-operators}

\begin{figure}[t]
  \centering
  \begin{subfigure}[b]{\linewidth}
    \centering
    \includegraphics[width=0.95\linewidth]{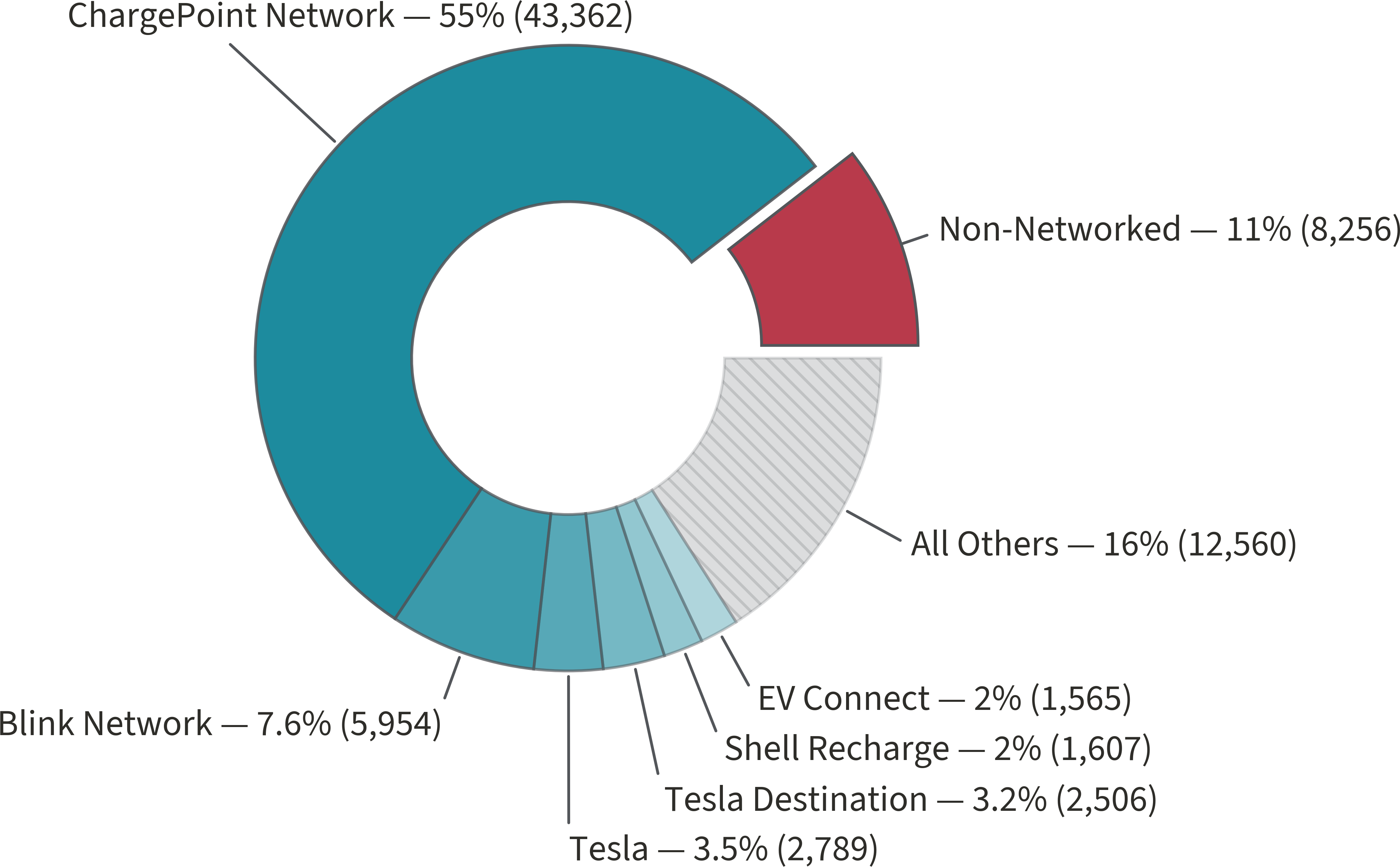}
    \caption{}
    \label{fig:public_networked_vs_non_networked}
  \end{subfigure}
  \hfill
  \vfill
  \begin{subfigure}[b]{\linewidth}
    \centering
    \includegraphics[width=0.95\linewidth]{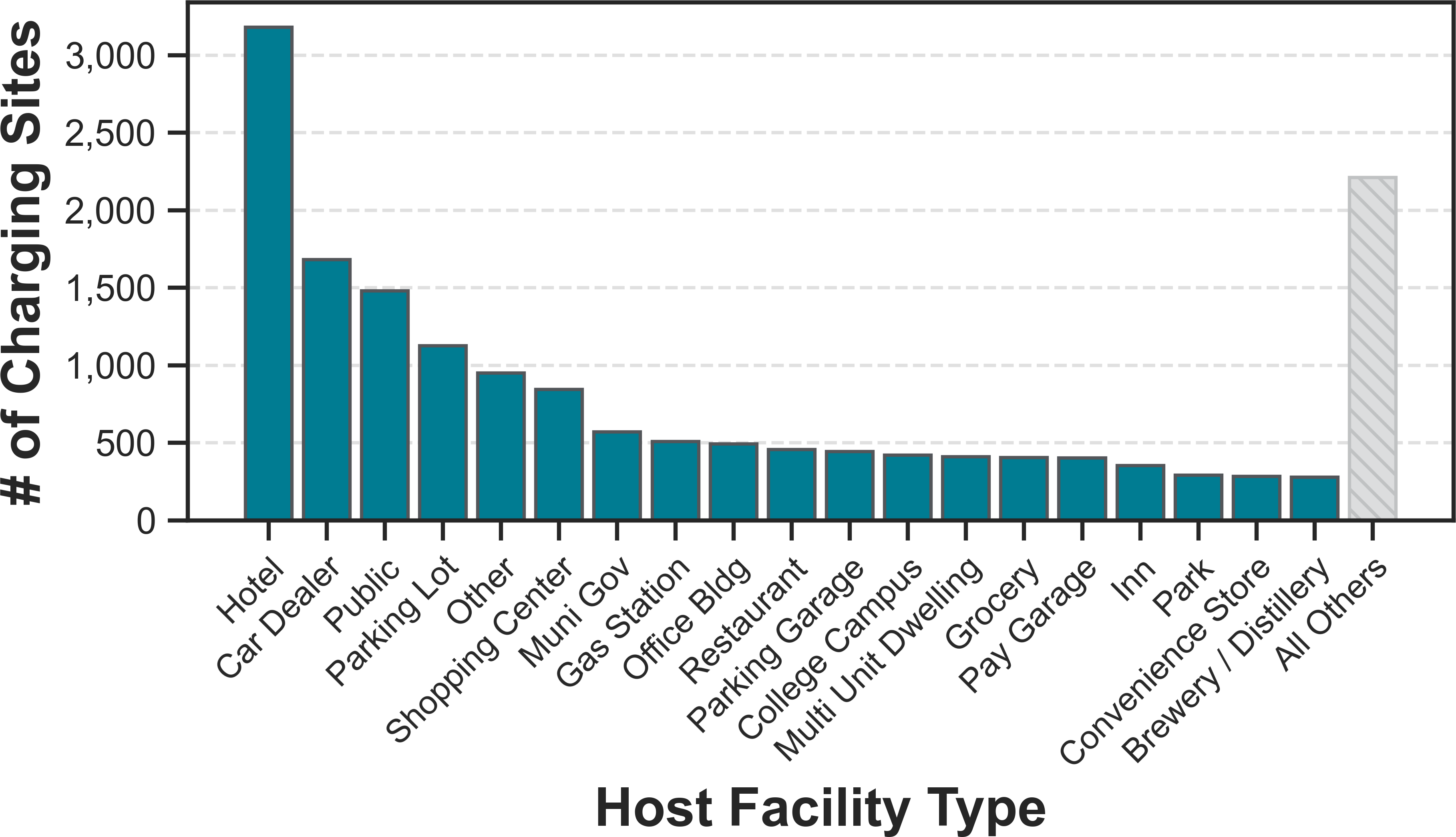}
    \caption{}
    \label{fig:public_facility_type}
  \end{subfigure}
  \caption{(a) Nearly 89\% of the 78,617 public charging sites in the U.S. are connected with a charging network; (b) public charging sites are commonly hosted by retail establishments, parking facilities, and workplaces\cite{u.s.departmentofenergyalternativefuelsdatacenterAlternativeFuelStations2025}.}
  \label{fig:pub}
\end{figure}

Public charging plays a crucial role in EV consumer confidence and meeting charging demand, especially for those without access to
reliable charging at home or work~\cite{wood2030NationalCharging2023}. Thus, public charging sites will remain a key
interface where several elements of charging infrastructure converge and interact. Charging sites physically host individual
charging stations as well as their associated electrical and local network infrastructure, and are where drivers connect
their vehicles and initiate charging. On-site chargers are often connected to a charging network operated by a
\emph{charging network provider} (CNP) that provides services to both end users and charging site operators, such as
user authentication, payment processing, installation, maintenance, and remote administration and monitoring. As of September 2025,
nearly 89\% of the 78,617 public U.S. charging sites are connected with a charging network
(Figure~\ref{fig:public_networked_vs_non_networked})~\cite{u.s.departmentofenergyalternativefuelsdatacenterAlternativeFuelStations2025}.

Overseeing the day-to-day operations of a charging site is the \emph{charging site operator} (CSO, or site operator),
which we define as the entity that ultimately has operational control and decision-making authority over all
charging-related aspects of a charging site, including charger installation, maintenance, pricing, and access
policies.\footnote{Site operators may subcontract third parties for some operational tasks, but retain ultimate
responsibility for the site.} Site operators can include dedicated chargepoint operators, universities, municipalities,
retail businesses, property management companies, or any other organization that hosts charging stations on their
premises.\footnote{While we describe the various stakeholders involved at charging sites as distinct entities for
  clarity, they may overlap in practice. For example, a site operator may also be a CNP (e.g., Tesla), or a CNP may also
be a charger manufacturer (e.g., ChargePoint).} Charging sites are
commonly hosted by retail establishments (e.g., hotels, shopping centers, grocery stores, restaurants),
parking lots and facilities, and workplaces (Figure~\ref{fig:public_facility_type})~\cite{u.s.departmentofenergyalternativefuelsdatacenterAlternativeFuelStations2025}.

Site operators are integral to \emph{assessing}, \emph{validating}, and \emph{addressing
} charging performance and
reliability at their sites. They are typically the first in line to hear from frustrated drivers when chargers are not
working as expected. However, while drivers and their experiences come and go, site operators remain constant, allowing
them to \emph{assess} whether or not charging issues reported by drivers are isolated incidents or part of larger, systematic
performance patterns that warrant further investigation. Towards this end, site operators typically have access to some degree of non-public charging
performance data provided by the CNP (or from the chargers themselves at non-networked sites) that they can use to \emph{validate} whether or not their site's reported performance matches the experiences of drivers and themselves. Finally, site operators ultimately decide whether or
not to take action to \emph{address} performance issues, which may include working with the CNP or charger manufacturer to validate,
troubleshoot, and fix issues.


Despite their pivotal role in assessing, validating, and addressing charging performance issues, site operators have been largely overlooked by existing efforts to improve U.S.
charging reliability in several key ways.

\subsection{Limited Access to Performance Information}
\label{sec:limited-data}

Standardization and recommendation efforts like the SAE CSPR Framework and the latest revisions of OCPP focus on improving the fidelity of performance data collected by charging stations and its reporting to a separate management system~\cite{CSPRFrameworkTechnical2023,WhatNewOCPP2023}. For networked chargers, this
management system is controlled by the CNP and not directly accessible by site operators. As nearly 89\% of U.S. public charging sites are networked (Figure~\ref{fig:public_networked_vs_non_networked}), the vast majority of site operators cannot directly access the
performance data reported from their site's chargers and must instead rely on their CNP to relay this information
to them.

However, there is no guarantee that CNPs will do so, or that the information shared
will be accurate, useful, or timely.
In fact, existing regulatory efforts like NEVI and those of the CEC do not enforce such data sharing. NEVI, for example, requires that
NEVI-funded charging sites must report some limited performance data to the federal government
and made available to the public, but do not explicitly require that site operators receive this information~\cite{NationalElectricVehicle,NationalElectricVehicle2023}.

Without guaranteed access to such performance information, site operators' ability to assess, validate, and address performance issues is drastically undermined. Specifically, simply validating whether or not a charger is experiencing performance issues becomes challenging, as site operators may be unable to conclusively corroborate or refute suspected performance issues. Additionally, such information is crucial for quantitatively evaluating the effectiveness of any remediation efforts made to address confirmed performance issues. Further still, site operators with access to such information may discover discrepancies or inaccuracies between their experiences and the reported performance information, which can be shared with CNPs and charger manufacturers to ultimately improve the accuracy and quality of performance reporting for all.

\subsection{Non-Actionable Performance Metrics for Site Operators}
\label{sec:limited-metrics}

Guaranteed access to performance information alone is not necessarily useful to site operators unless that information directly helps them assess, validate, and address performance issues at their sites.

Given the prevalence and improved performance reporting capabilities of OCPP~\cite{WhatNewOCPP2023}, it is tempting to simply provide site operators access to all of the performance data emitted directly from their chargers. However, this information may be too technical, verbose, and fine grained to be immediately useful to site operators, especially considering that most charging sites are hosted by retail or other establishments that may not possess the expertise required to process and interpret this data.
Instead, site operators require performance metrics that are readily interpretable and actionable without complex data processing or an intimate understanding of the underlying charging hardware. 

At first glance, the uptime metric provided by NEVI (and adopted by the CEC~\cite{adamdavisAssemblyBill21272024,robinsonSecondDraftStaff}) appears to fit this criteria. Per NEVI, a charger is considered ``up'' when its hardware and software are both online and available for use, or in 
use, and successfully 
dispenses electricity in accordance with requirements for minimum power level, and is defined formally as follows:

\begin{equation*}
\mu = \frac{525,600 - \left(T_\text{outage} - T_\text{excluded}\right)}{525,600} \times 100, 
\end{equation*}
where \(\mu\) denotes charging station\footnote{As a single charging station may have multiple charging ports, NEVI uptime refers to that of individual charging ports. However, for clarity and consistency, we simplify this to refer to the station only.} uptime as a percentage of the previous year, \(T_\text{outage}\) denotes the total minutes of outage in the previous year, and \(T_\text{excluded}\) denotes the total minutes of outage in the previous year caused by reasons outside the site operator's control~\cite{NationalElectricVehicle2023}.

However, NEVI uptime has been criticized as ``relatively simplistic'' and ``necessary but not
sufficient''~\cite{jeffst.johnWhatsEpidemicUnreliable2023} to ensure charging reliability at large. Additionally, we highlight that this metric specifically undermines site operators' ability to assess, validate, and address performance issues. For one, uptime alone is opaque and non-actionable for site operators in that it does not provide any reasoning as to \emph{why} a charger is not ``up''. Without such reasoning, site operators are left in the frustrating position of knowing that their chargers are experiencing issues while lacking the information necessary to effectively address them. Secondly, uptime is defined only for year- and month-long time periods, meaning that site operators must wait at least that long for the metric to be available to them. Such long reporting periods effectively delay site operators' ability to timely assess, validate, and address issues. Additionally, such long reporting periods can conceal finer patterns of performance issues that occur on shorter intervals (e.g., weekly, daily, minutely, etc.).

%% file: sections/4-new-metrics.tex
\section{Actionable Performance Metrics for Site Operators}\label{sec:new-metrics}

In summary, limited access to performance information and non-actionable metrics undermine site operators' ability to assess, validate, and address performance issues at their sites.

Recognizing these limitations, we propose a set of performance metrics that address the shortcomings detailed in the previous section. Namely, our metrics directly improve upon NEVI uptime by providing site operators with insights into \emph{why} their sites' chargers may not be working as expected, thereby enabling appropriate action. Furthermore, our metrics are defined over a configurable period of interest, allowing them to be evaluated at annual, monthly, weekly, daily, or any other frequency desired.
Lastly, while our metrics do not directly improve site operators' access to performance information on their own, we posit---and demonstrate in Section~\ref{sec:case-study}---that they are readily computable without requiring new standards or changes to existing chargers; thus, with some regulatory pressure, they could be made available to many site operators nationwide.

Our metrics decompose charger behavior into simple yet operationally meaningful states defined as follows:
\begin{itemize}
\item \textsc{\textbf{up}}: The charger is ``up'' per the NEVI definition provided in Section~\ref{sec:limited-metrics}.
\item \textsc{\textbf{faulted}}: The charger is in a self-reported nonfunctional, malfunctioning, or faulted state.
\item \textsc{\textbf{unreachable}}: The charger has lost network connectivity and cannot communicate with its central management system.
\item \textsc{\textbf{unavailable}}: The charger has been explicitly powered down or otherwise taken offline by the site operator.
\end{itemize}

Each of our metrics is described as follows, where \emph{i)} \(C\) denotes the charger under analysis; \emph{ii)} \(P\) and \(T_{P}\)
denote the time period of interest and the duration of that period, respectively; and \emph{iii)} \(T_{C, P, s}\) denotes the duration charger \(C\) spent in state \(s\nolinebreak\in\nolinebreak\{\textsc{\textbf{up}}, \textsc{\textbf{faulted}}, \textsc{\textbf{unreachable}}, \textsc{\textbf{unavailable}}\}\) during period \(P\).

\subsection{Uptime}
\emph{Uptime} is a straight-forward extension of NEVI uptime that can be evaluated over arbitrary time windows. Uptime is defined as the percentage of time that a charger is in the \textsc{\textbf{up}} state over the period of interest and is calculated as follows:\footnote{While we omit an analog of NEVI excluded time in our Uptime definition for clarity, it can easily be extended with such.}

\begin{equation}
\text{Uptime}(C, P) = \frac{T_{C, P, \textsc{\textbf{up}}}}{T_{P}} \times 100
\end{equation}

\subsection{Fault Time \& Fault-Reason Time}

\emph{Fault Time} describes periods of time when a charger is down due to self-reported internal errors or malfunctions that typically require maintenance intervention. Fault Time is defined as the percentage of time that a charger is in the \textsc{\textbf{faulted}} state over the period of interest and is calculated as follows:

\begin{equation}
\text{Fault Time}(C, P) = \frac{T_{C, P, \textsc{\textbf{faulted}}}}{T_{P}} \times 100
\end{equation}

\emph{Fault-Reason Time} breaks down Fault Time by reasons reported by the charger or CNP,\footnote{Standardization and accuracy of reported reasons is beyond the scope of this work. Refer to~\cite{mariaeduardamontezzocoelhoRecommendationsMinimumRequired2025} for recommendations regarding minimum required error codes.} allowing for focused diagnosis and prioritization of corrective measures. Fault-Reason Time is defined as the percentage of time that a charger is faulted due to a specific reason over the period of interest and is calculated as follows:

\begin{equation}
\text{Fault-Reason Time}(C, P, R) = \frac{T_{C, P, \textsc{\textbf{faulted}}, R}}{T_{C, P, \textsc{\textbf{faulted}}}} \times 100,
\end{equation}
where \(T_{C, P, \textsc{\textbf{faulted}}, R}\) is the duration that charger \(C\) was faulted due to reported reason \(R\) during period \(P\).


\subsection{Unreachable Time}

\emph{Unreachable Time} captures loss of communication between a charger and its management system, which can prevent remote
monitoring and control, and may interrupt charging sessions due to failed user authentication or payment authorization. Unreachable time is defined as the percentage of time that a charger is in the \textsc{\textbf{unreachable}} state
over the period of interest and is calculated as follows:

\begin{equation}
\text{Unreachable Time}(C, P) = \frac{T_{C, P, \textsc{\textbf{unreachable}}}}{T_{P}} \times 100
\end{equation}

%% file: sections/5-case-study.tex
\section{Case Study}
\label{sec:case-study}

In this section, we present a case study to demonstrate that our proposed performance metrics are able to provide actionable insights to site operators, yet are simple enough to be readily computable at existing charging sites.

The charging site under study is hosted by an academic institution in California, and consists of 98 6-kW AC Level 2 chargers ranging between 1 and 11 years old, with a mean age of approximately 7.5 years. All on-site chargers are networked with a prominent international charging network that serves over half of all public
charging sites in the U.S.

We compute our metrics for this site using \evsee{}, our open-source charging site performance reporting tool. \evsee{} is designed to compute our metrics using only data that is already available from existing chargers and charging networks; in other words, \evsee{} does not require any modifications to existing chargers or downstream systems.

In summary, \evsee{} performs three operations. First, it extracts raw charger data from preexisting external source systems (e.g., charging network dashboards, management systems, or individual chargers). Second, it normalizes this data into a consistent format that captures charger metadata, operational status, faults, and charging sessions. Finally, it calculates our performance metrics from the normalized data and presents them to site operators through a user-friendly web interface. \evsee{} is described in detail in Appendix~\ref{sec:evsee}.

\evsee{} is made compatible with existing systems using modular \emph{integration
plugins}, which essentially provide the data extraction and data normalization implementation for a given source system. In this case study, the source system is the CNP-provided charger administration web dashboard for the charging site under study. This dashboard contains real-time operational status information and historical event logs (including faults), which was extracted and normalized by \evsee{} to compute each metric at yearly, monthly, and daily intervals for each on-site charger active during the period of interest. Altogether, the computed metrics represent approximately seven years of site operation from 2018 through 2024. The full details of the \evsee{} integration module developed for this case study can be found in Appendix~\ref{sec:integration-plugin}.

\begin{figure}[t]
  \centering
  \begin{subfigure}[b]{\linewidth}
    \centering
    \includegraphics[width=0.95\linewidth]{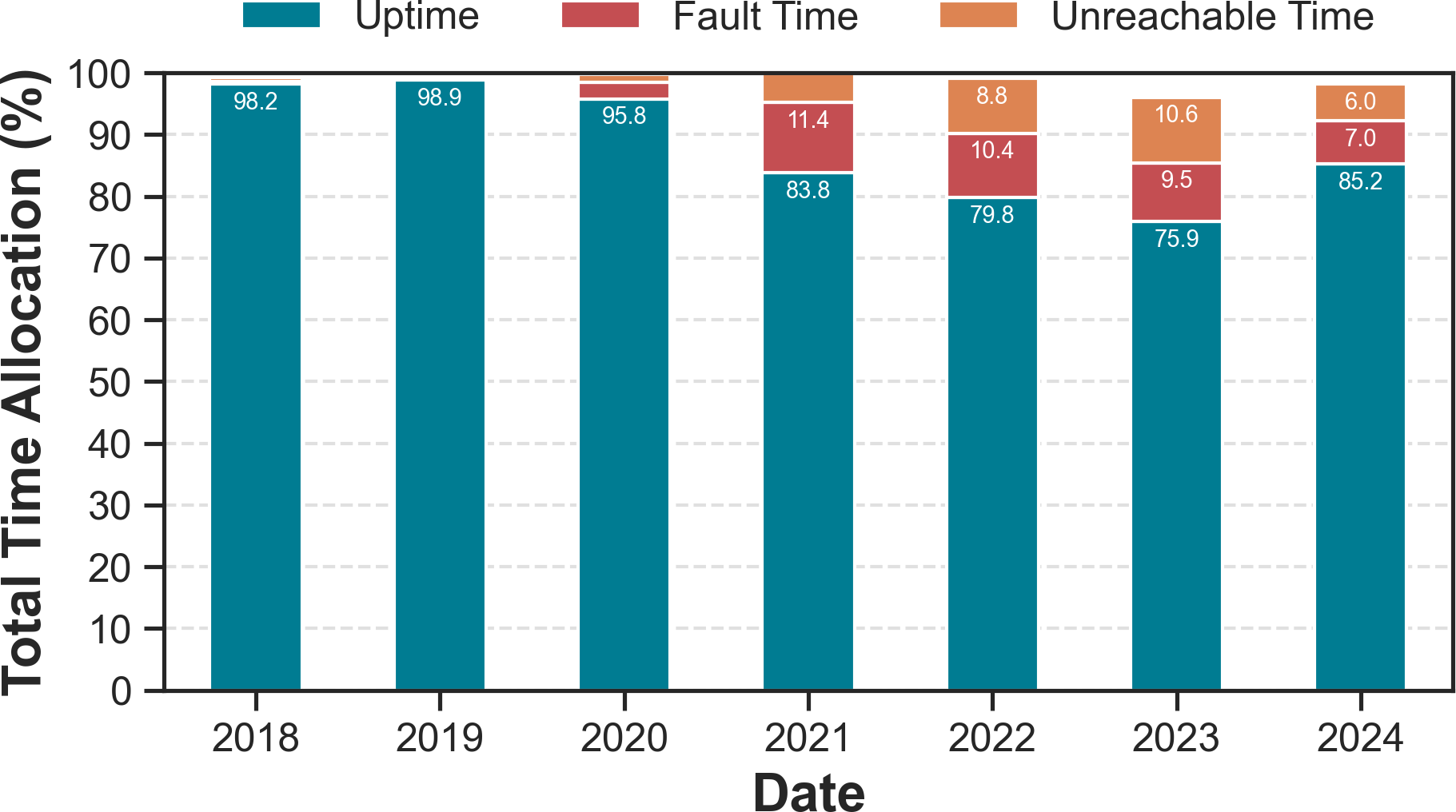}
    \caption{}
    \label{fig:yearly_avg}
  \end{subfigure}
  \hfill
  \vfill
  \begin{subfigure}[b]{\linewidth}
    \centering
    \includegraphics[width=0.95\linewidth]{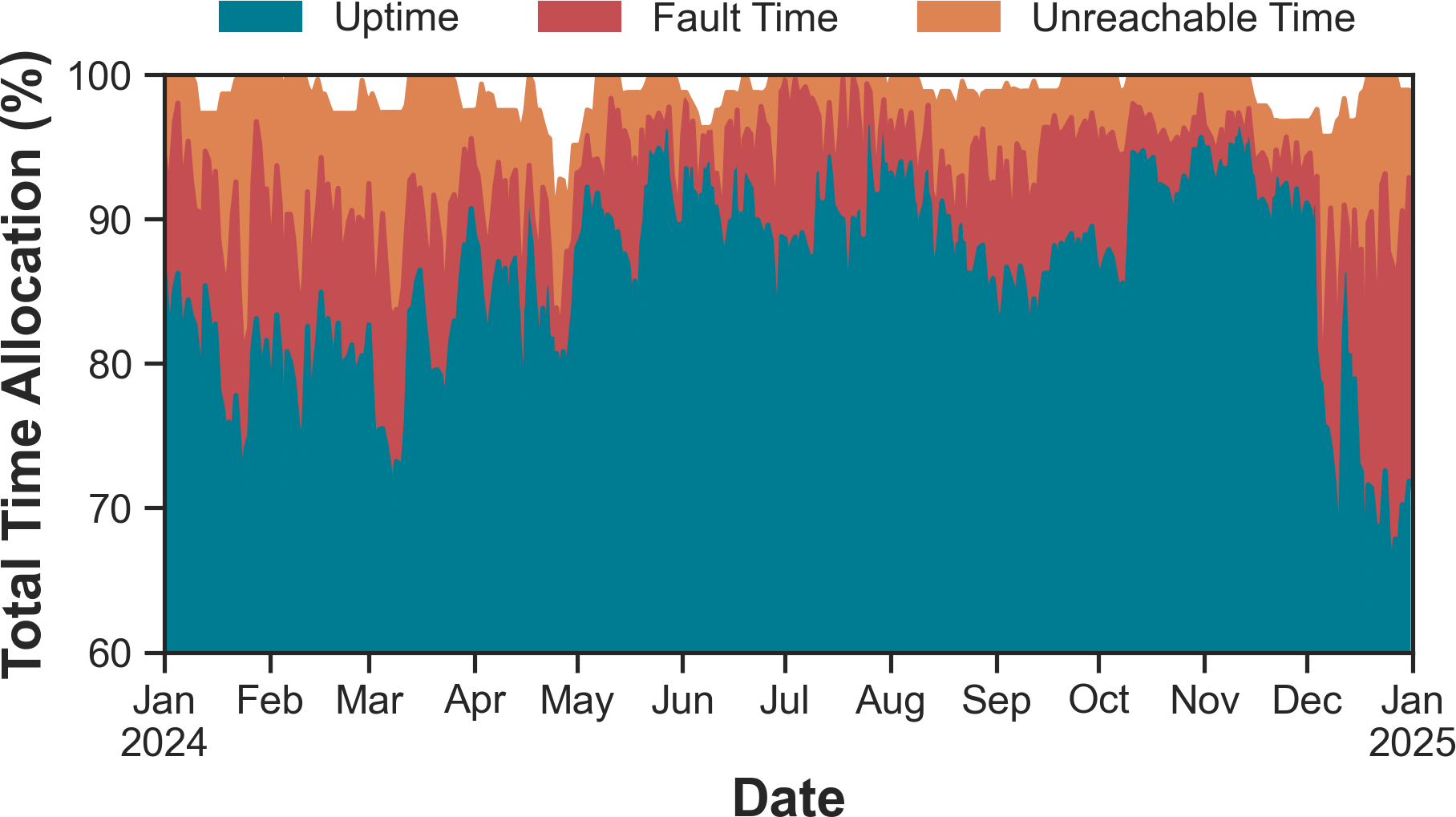}
    \caption{}
    \label{fig:daily_avg}
  \end{subfigure}
  \caption{Decomposed site-wide average Uptime, Fault Time, and Unreachable Time for the case study site. (a) Yearly averages (2018--2024) illustrate how these metrics discriminate between causes of downtime, such as self-reported faults peaking in 2021 versus network-driven failures becoming the primary driver by 2023. (b) Daily averages (2024) demonstrate the actionable benefit of high-resolution reporting, revealing frequent connectivity spikes and persistent ``zombie'' chargers that remain hidden within coarser yearly averages.}
  \label{fig:avg}
\end{figure}

\begin{figure}[t]
  \centering
  \begin{subfigure}[b]{\linewidth}
    \centering
    \includegraphics[width=0.95\linewidth]{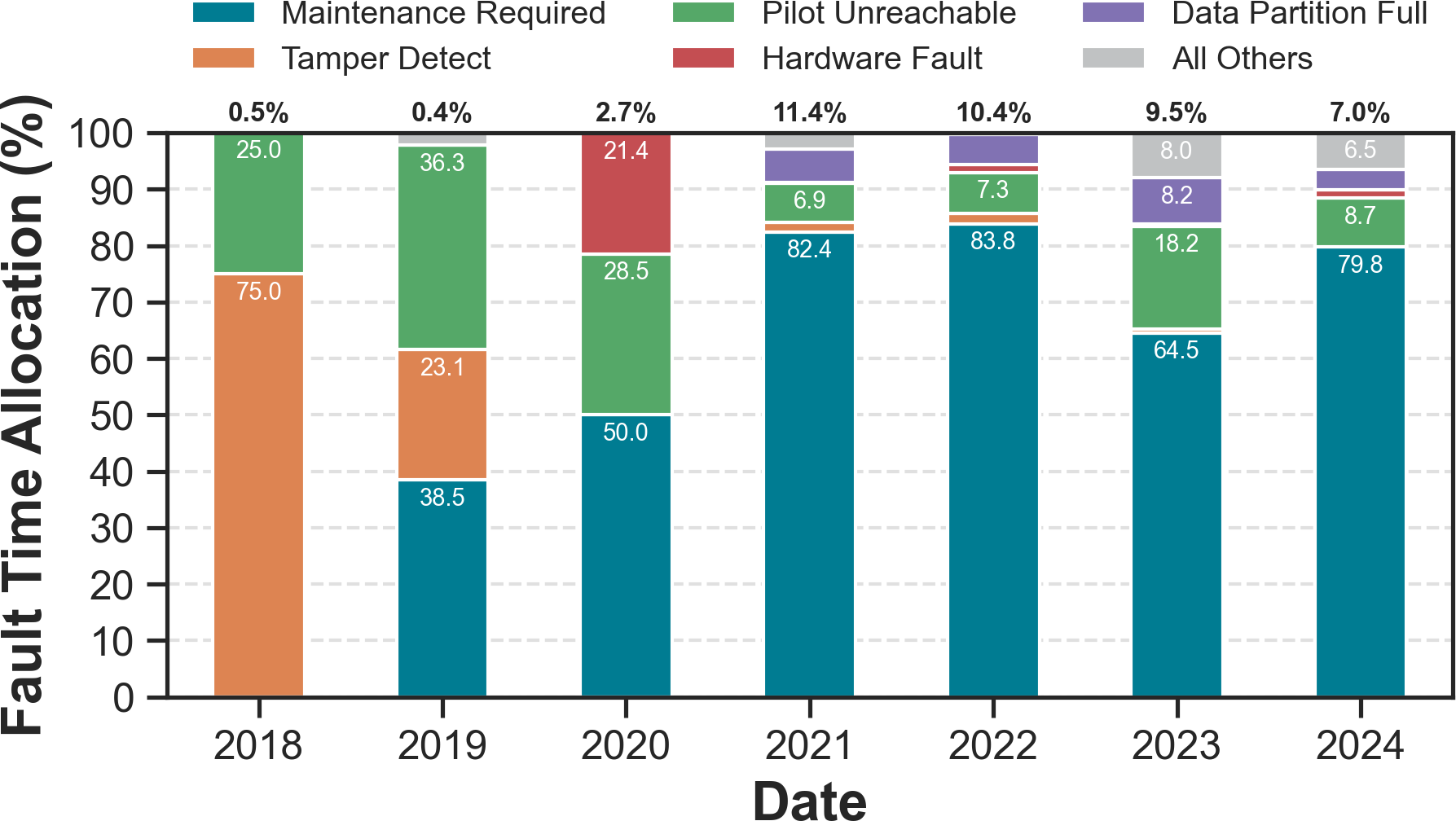}
    \caption{}
    \label{fig:sft_yearly}
  \end{subfigure}
  \hfill
  \vfill
  \begin{subfigure}[b]{\linewidth}
    \centering
    \includegraphics[width=0.95\linewidth]{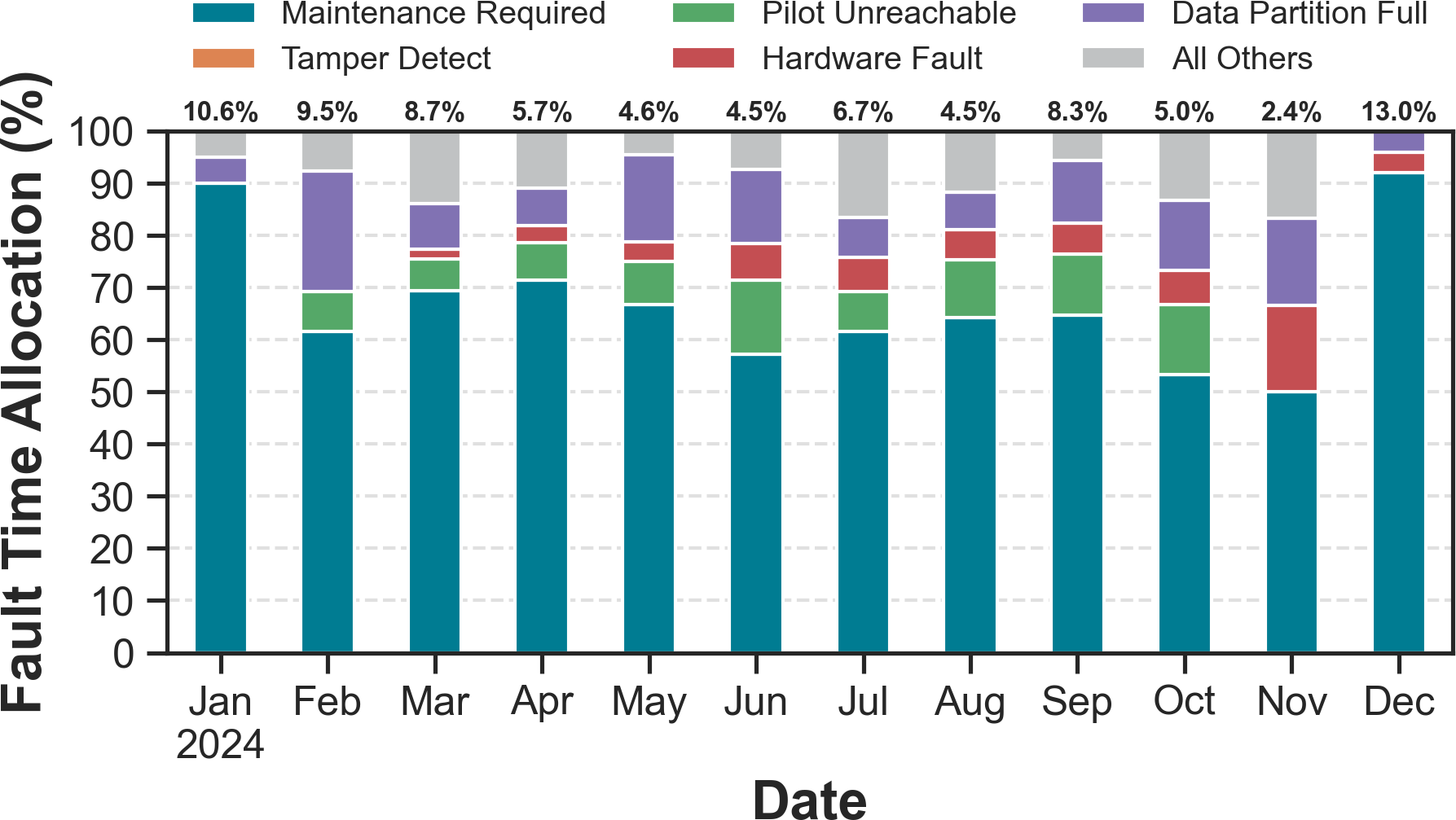}
    \caption{}
    \label{fig:sft_monthly}
  \end{subfigure}
  \caption{Decomposed site-wide average Fault-Reason Time for the case study site. (a) Yearly averages (2018–2024) illustrate how self-reported faults evolved from early physical ``Tamper Detect'' events to systemic hardware malfunctions and firmware-driven ``Data Partition Full'' errors over time. (b) Monthly averages (2024) provide actionable diagnostic evidence by identifying a persistent baseline of ``Maintenance Required'' faults across the fleet.}
  \label{fig:sft}
\end{figure}

Figure~\ref{fig:avg} depicts site-wide average Uptime, Fault Time, and Unreachable Time across two views: (a) yearly averages from 2018--2024 and (b) daily averages for 2024.

As revealed in Figure~\ref{fig:yearly_avg}, the site maintained an exceptional level of performance during 2018 and 2019, with Uptime exceeding 98\% and negligible interruptions from faults or connectivity issues. However, starting in 2020, the data reveals a clear and steady degradation in performance, with Uptime reaching a low of 75.9\% in 2023. Notably, Fault Time and Unreachable Time provide crucial insight into why and how performance degraded over time. For instance, while Fault Time (red) peaked in 2021 at 11.4\%, indicating a period of significant self-reported hardware malfunctions, Unreachable Time (orange) became the primary driver of downtime in 2023, accounting for 10.6\% of total time. Compared to Uptime alone, our Fault Time and Unreachable Time metrics provide a critical diagnostic distinction, revealing whether local networking infrastructure—rather than the chargers themselves—requires attention. Even without granular fault reasons, this differentiation provides immediate actionable value by empowering site operators to move beyond vague complaints and initiate targeted, data-backed discussions with charging network providers, manufacturers, or internal IT teams to address the root causes of poor performance.

While the yearly averages provide a macro-level assessment of site performance, they inherently mask the day-to-day volatility that defines the actual experience of EV drivers, and consequently, site operators. As shown in Figure~\ref{fig:daily_avg}, the fine-grained daily resolution reveals significant daily inconsistency in charger availability despite the relatively improved yearly Uptime of 85.2\%. Performance is characterized by frequent, sharp spikes in Unreachable Time, suggesting intermittent local network instability that would be invisible in coarser, year-end reporting. Most strikingly, the daily view captures a major performance drop in late December 2024, where Uptime fell below 70\% for several days.
Additionally, the steady baseline of Fault
Time and Unreachable Time throughout the year implies a group of ``zombie''
chargers with persistent hardware and networks issues that require further investigation.

Similarly, Figure~\ref{fig:sft} depicts site-wide average Fault-Reason Time across two views: (a) yearly averages from 2018–2024 and (b) monthly averages for 2024.

As shown in Figure~\ref{fig:sft_yearly}, the fault reasons at the case study site evolved significantly over time. In earlier years (2018–2019), the negligible fault time was dominated by ``Tamper Detect'' events, suggesting that external physical interactions or sensor sensitivities---rather than internal malfunctions---were the primary concern. However, by 2020, there was a distinct shift toward ``Hardware Fault'' and ``Maintenance Required'' categories, potentially signaling the onset of mechanical or electrical wear as the units aged. Notably, the emergence of the ``Data Partition Full'' error in 2021 points to a systemic firmware or software issue that remained a persistent baseline problem through 2024.

The monthly resolution provided in Figure~\ref{fig:sft_monthly} further highlights the prevalence of ``zombie'' chargers that remain unresolved over long periods. Throughout 2024, ``Maintenance Required'' remained the dominant fault category, consistently accounting for a significant portion of the Fault Time allocation in any given month. This steady baseline of maintenance needs is punctuated by intermittent spikes in other failure modes, such as a pronounced increase in ``Hardware Fault'' allocation during November 2024.


%% file: sections/6-conclusion.tex
\section{Conclusion}\label{sec:conclusion}

In summary, this paper highlights site operators' crucial role in assessing, validating, and addressing charging performance and reliability issues in the U.S., and how this role has been overlooked by existing performance improvement and reporting efforts.  Specifically, we scrutinize site operators' limited access to useful performance data and how existing, non-actionable performance metrics fail to meet their needs. In response, we propose a set of readily computable, actionable performance metrics for site operators---namely Fault Time, Fault-Reason Time, and Unreachable Time---that address such shortcomings. We demonstrate through a case study that our proposed metrics are not only readily computable at existing charging sites, but capable of deriving actionable intelligence at the diagnostic granularity required for site operators to effectively assess, validate, and ultimately address performance issues at their sites.

%% file: sections/A-evsee.tex
\section{\evsee{} Explained}
\label{sec:evsee}

\evsee{} derives charger performance metrics from charger data extracted, consolidated, and normalized from
pre-existing external systems (e.g., charging network-provided APIs or web dashboards, charger management systems, or
individual charging units). This data and derived metrics are then made available to site operators via a user-friendly,
interactive web frontend. Crucially, \evsee{}'s approach does not require any modifications to the chargers themselves
or any downstream systems, and can thus be applied to existing charging sites. To the best of our knowledge, \evsee{} is
the first publicly available tool that provides performance reporting to site operators in this manner.

\evsee{} consists of the main \evsee{} Python application, \emph{integration plugins} that provide compatibility with
various external systems, an interactive web interface powered by Apache Superset, and
supporting PostgreSQL relational database instances. The main \evsee{} Python application
configures and orchestrates \evsee{}'s three core operations: \emph{i)} data extraction, \emph{ii)} data normalization,
and \emph{iii)} performance metrics calculation. \evsee{}'s entire application stack is containerized using Docker
Compose, allowing it to be easily deployed and configured either on premises or in the cloud.

The following subsections describe \evsee{}'s core operations and components in more detail.



\subsection{Data Extraction}

In this operation, raw charger data from pre-existing external systems is extracted and consolidated into \evsee{}'s raw
data table. External systems refer to any underlying data source that provides information about a site's chargers, such
as charging network-provided APIs or web dashboards, charger management systems, or individual charging units (via OCPP
or another protocol). While available data and their formats vary across external systems, they typically include charger metadata, status, faults, and charging sessions.

The exact data extraction implementation for a given external system is provided by its associated \emph{integration
plugin}. These integration plugins are modular, allowing \evsee{} to be easily extended to support new external systems
without modifying the main \evsee{} application. In Appendix~\ref{sec:integration-plugin}, we describe
the development of the integration plugin used for \evsee{}'s evaluation in detail.

Extracted raw data is consolidated into the raw data table.
Each entry in the raw data table includes a unique identifier for the raw data item, a timestamp indicating when the
data was extracted, a boolean flag indicating whether the data has been processed, and the raw data item itself in a
format determined by the integration plugin. Additionally, a field is included to specify the data type of the raw data
item, which is used later by the integration plugin to determine how to process the item during data normalization.

This approach provides two key benefits: \emph{i)} it provides a consistent retrieval interface for all raw data,
regardless of its source; and \emph{ii)} it provides a comprehensive historical record of raw charger data, which may
not be available from the source system or otherwise accessible by site operators.

\subsection{Data Normalization}
\label{sub-sec:data-normalization}

In this operation, unprocessed external system-specific raw charger data is retrieved from the raw data store in
chronological order and transformed into \evsee{}'s normalized charger information model. As with data extraction, the
exact procedure for this transformation is defined by the integration plugin for the specific external system. The
normalized model charger information model is composed of the following data tables.

\subsubsection{Charger Metadata}

This table contains identifying information about all known chargers. At minimum, each entry includes the manufacturer,
serial number, and location for each charger, but may also contain additional information such as model, power rating,
and installation date if available. Additionally, each charger is assigned a unique identifier that is used to reference
it in other tables.

\subsubsection{Charger Status}\label{sub-sub-sec:charger-status}

This table contains information about the status of a charger over time. Each entry includes the unique identifier of
the charger, a timestamp indicating when the status was recorded, and the charger's status at that time. \evsee{}
assumes that a charger can be in one of the following six states at any given moment:

\begin{itemize}
\item \textsc{\textbf{occupied}}: the charger is currently in use.
\item \textsc{\textbf{available}}: the charger is not currently in use but is otherwise functional.
\item \textsc{\textbf{unavailable}}: the charger has been explicitly taken out of service or powered down by the site operator.
\item \textsc{\textbf{faulted}}: the charger is malfunctioning.
\item \textsc{\textbf{unreachable}}: the charger cannot be reached or contacted.
\item \textsc{\textbf{unknown}}: the charger's status is not known.
\end{itemize}

\subsubsection{Charger Faults}

This table contains information about faults or issues reported by a charger. Each entry includes the unique identifier
of the charger, a timestamp indicating when the fault was reported, and a textual name or description of the fault. As
faults and issues can vary widely across different chargers and systems, \evsee{} does not impose any specific
standardization on the fault names or descriptions.

There is an implicit relationship between the \textsc{\textbf{faulted}} state in the charger status table and the
charger faults table: if a charger is in the \textsc{\textbf{faulted}} state at a given time, it is assumed that the
last fault reported in the charger faults table at or before that time is the cause of the faulted state.

\subsubsection{Charging Sessions}\label{sub-sub-sec:charging-sessions}

This table contains information about charging sessions that have occurred at a charger. Each entry includes the unique
identifier of the charger, a timestamp indicating when the session started, a timestamp indicating when the session
ended, and the total energy consumed during the session.

\subsection{Performance Metrics Calculation}

As its name suggests, this operation calculates performance metrics from the normalized charger information model.
\evsee{} calculates performance metrics for each charger over several time scales: daily, weekly, monthly, and yearly following the formulas described in Section~\ref{sec:new-metrics}.

\subsection{Configuration \& Orchestration}

The aforementioned operations are orchestrated by the main \evsee{} Python application according to a user-defined
configuration file. This configuration files describes \emph{i)} which external systems to extract data from and the
settings (e.g., API keys, login credentials, URLs, etc.) of their respective integration plugins; and \emph{ii)} the
frequency at which data extraction and normalization should occur (e.g., hourly, daily, etc.). The configuration file is
written in YAML, which is human-readable and easy to modify.

\subsection{Web Interface}

The web interface is powered by Apache Superset, which provides an interactive data visualization and exploration
platform. All of the normalized charger information model data and derived performance metrics are made available to
site operators via this web interface, which allows them to view and analyze charger performance metrics in a
user-friendly manner. The web interface provides various visualization options, such as time series graphs, bar charts,
and tables. The web interface also allows site operators to filter and group data by various attributes (e.g., charger
location, time period, etc.) to gain insights into charger performance across different dimensions, as well as view
detailed information about individual chargers, configure personalized dashboards, and export data for further analysis.

%% file: sections/B-integration-plugin.tex
\section{\evsee{} Integration Plugin for Case Study}
\label{sec:integration-plugin}

As described previously, \evsee{} relies on integration plugins to provide the data extraction and normalization implementations for a specific external system. In this section, we describe the integration plugin developed for the charging network used at the charging site evaluated in the case study presented in Section~\ref{sec:case-study}, henceforth referred to as the \emph{source system}.

\subsection{Data Extraction}

The data extraction component of the integration plugin extracts raw data from the source system. The source system
provides a web-based administrative dashboard for the site operator to manage and monitor the site's chargers. As the
source system does not provide an API or any other programmatic access method, the integration plugin uses a browser
automation-based web scraping method to extract data from the dashboard.

The integration plugin extracts information from three areas of the dashboard: \emph{i)} the charger status overview
table; \emph{ii)} the charger events table; and \emph{iii)} the charging sessions table. Notably, the source system does
\emph{not} directly provide any performance metrics (e.g., uptime, etc.).

The charger status overview table contains summary information for each charger at the site. Each entry includes the
charger's serial number, address, geographic coordinates, and current status. The statuses reported by the source system
are analogous to those defined in \evsee{}'s normalized charger information model described in
Appendix~\ref{sub-sub-sec:charger-status} (excluding the \textsc{\textbf{unknown}} state). Crucially, only the
\emph{current} (i.e., at the time of dashboard access) status is provided; that is, the source system does \emph{not}
provide historical status information for each charger. From this table, the integration plugin creates a single raw
data item of type \texttt{StationOverview} consisting of all entries in the table at the time of extraction.

The charger events table logs events of interest for each charger at the site. Each entry includes the charger's serial
number, a timestamp indicating when the event occurred, and the event's name. As presented by the source system, these
event entries are not useful: they do not include any additional information about the event, such as its severity or
impact on the charger's status. However, as we will describe shortly, they can be used to infer the charger's status at
the time of the event. The integration plugin creates a raw data item of type \texttt{ChargerEvent} for each entry in
the table.

Lastly, the charging sessions table describes all charging sessions that have occurred at the site in a format similar
to that used in \evsee{}'s normalized information model described in Appendix~\ref{sub-sub-sec:charging-sessions}. The
integration plugin creates three raw data items for each entry in the table: an item of type \texttt{ChargingSession}
for the entry itself; and two additional items for the start and end events of the charging session of types
\texttt{ChargingSessionStart} and \texttt{ChargingSessionEnd}, respectively. These events will be used to infer when the
charger is occupied in the subsequent data normalization step.

\subsection{Data Normalization}

The data normalization component of the integration plugin updates \evsee{}'s normalized charger information model
according to the type of the raw data item being processed as follows:

\subsubsection*{\texttt{StationOverview}}

New entries in the Charger Metadata table are created from the serial numbers in the raw data item if they do not
already exist. Then, the current status of each charger is updated in the Charger Status table according to the current
status and extraction timestamp reported in the raw data item.

\subsubsection*{\texttt{ChargerEvent}}

A new entry in the Charger Metadata table is created from the serial number in the raw data item if it does not already
exist. The event is then used to infer the charger's status at the time of the event. While the source system does not
explicitly provide the status of the charger at the time of the event, some event names are clearly indicative of the
charger's status. Furthermore, several event names and their meanings are documented in the charger manufacturer's
operation and maintenance manual. Using these resources, as well as experimentation with the source system, we classify
the events into the following categories:

\begin{itemize}
\item Informational
\item Faults
\item Fault Cleared
\item Network
\item Power
\end{itemize}

From these categories, we developed a finite state machine-based method that is used to predict
the charger's status at the time of the event based on the event's category and the current status of the corresponding
charger. This predicted status is used to create a new entry in the Charger Status table for the corresponding charger
at the time of the event.


\subsubsection*{\texttt{ChargingSession}, \texttt{ChargingSessionStart}, and \texttt{ChargingSessionEnd}}

A new entry in the Charger Metadata table is created from the serial number in the raw data item if it does not already
exist. For \texttt{ChargingSession} items, a new entry is created in the Charging Sessions table corresponding to the
item. For \texttt{ChargingSessionStart} items, a new entry is created in the Charger Status table with state
\textsc{\textbf{occupied}} for the corresponding charger. Similarly, for \texttt{ChargingSessionEnd} items, a new entry
is created in the Charger Status table with state \textsc{\textbf{available}} for the corresponding charger.